\documentclass[12pt]{article} 
\usepackage{amstex}
\usepackage{amssymb}
\usepackage{citesort}
\usepackage{graphicx}                              
\usepackage[nomarkers]{endfloat}
\usepackage{pstricks}
\numberwithin{equation}{section}
\begin{document}

\setcounter{page}{0}
\thispagestyle{empty}

\begin{flushright}
{\small BARI-TH 324/99}
\end{flushright}

\vspace*{2.5cm}

\begin{center}
{\large\bf Probing the non-perturbative dynamics of SU(2) vacuum}
\end{center}

\vspace*{2cm}

\renewcommand{\thefootnote}{\fnsymbol{footnote}}

\begin{center}
{
Paolo Cea$^{1,2,}$\protect\footnote{Electronic address:
cea@@bari.infn.it} and
Leonardo Cosmai$^{2,}$\protect\footnote{Electronic address:
cosmai@@bari.infn.it} \\[0.5cm]
$^1${\em Dipartimento di Fisica, Universit\`a di Bari,
I-70126 Bari, Italy}\\[0.3cm]
$^2${\em INFN - Sezione di Bari,
I-70126 Bari, Italy}
}
\end{center}

\vspace*{0.5cm}

\begin{center}
{
February, 1999
}
\end{center}

\vspace*{1.0cm}

\renewcommand{\abstractname}{\normalsize Abstract}
\begin{abstract}
The vacuum dynamics of SU(2) lattice gauge theory is studied by means of a gauge-invariant
effective action defined using the lattice Schr\"odinger functional. 
Numerical simulations are performed both at zero and finite temperature.
The vacuum is probed using an external constant Abelian chromomagnetic field. 
The results
suggest that at zero temperature the external field is screened in the continuum limit.
On the other hand at finite temperature it seems that confinement is restored by
increasing the strength of the applied field.
\end{abstract}

\vspace*{0.5cm}
\begin{flushleft}
PACS number(s): 11.15.Ha
\end{flushleft}

\renewcommand{\thesection}{\normalsize{\Roman{section}.}}
\section{\normalsize{INTRODUCTION}}
\renewcommand{\thesection}{\arabic{section}}

It is widely recognized that the effective action is a useful tool to 
investigate the quantum properties of field theories.

In the case of gauge theories when including the quantum fluctuations 
one faces  the problem to retain in the effective action the gauge invariance 
that is manifest at the classical level. In the perturbative approach, however, 
the problem of the gauge invariance of the effective action is not so 
compelling. Indeed in order to perform the perturbative calculations we need to 
fix the gauge so that the gauge invariance is lost anyway. Obviously the
physical quantities turn out to be gauge invariant. In the case of the 
perturbative evaluation of the effective action the problem of the gauge
invariance can be efficiently resolved by the so-called method
of the background effective action~\cite{Schwinger1951,Honerkamp,tHooft}. 
In the background field approach one separates the quantum field into
the fluctuations $\eta(x)$ and the background field $\bar{A^a_\mu}(x)$.
In order to define the background field effective action we introduce the 
partition function by coupling the external current to the fluctuations.
Using the background field gauge fixing it is easy to see that 
the partition function is invariant against gauge transformation of the
background field. In this way, after performing the usual Legendre transformation,
one obtains an effective action which is invariant for background field 
gauge transformations. 

The lattice approach to gauge theories allows
the non perturbative study of gauge systems without loosing the gauge invariance.
Thus, it is natural to seek for a lattice definition of the effective action.
Previous attempts (both in three~\cite{Cea3dim,Trottier} 
and four~\cite{Cea4dim,Ambjorn89,Levi}
dimensions) in this direction introduced the background field by means of an 
external current coupled to the lattice gauge field.  
It turns out, however, that the current term added to the lattice gauge action
in general is not invariant under the local gauge transformations belonging to the 
gauge group. For instance, if one considers abelian background fields then the
action with the current term turns out to be invariant only for 
an abelian subgroup of the gauge group. So that only in the case of
abelian U(1) gauge theory the latttice background field action is gauge invariant.

The aim of the present paper is to discuss in details a recently proposed
method~\cite{metodo} to define on the lattice the gauge invariant effective
action by using the so-called 
Schr\"odinger functional~\cite{Rossi1980,Gross1981,Luescher}.

Let us consider the continuum Euclidean Schr\"odinger functional in Yang-Mills
theories without matter field:
\begin{equation}
\label{Zeta}
{\mathcal{Z}}[A^{(f)},A^{(i)}] = 
\left\langle  A^{(f)} \left| e^{-HT} {\mathcal{P}} \right| A^{(i)} \right\rangle \,.
\end{equation}
In Eq.~(\ref{Zeta}) $H$ is the pure gauge Yang-Mills Hamiltonian in the
fixed-time temporal gauge, $T$ is the Euclidean time extension, while
${\mathcal{P}}$ projects onto the physical states. $A^{a(i)}_k(\vec{x})$
and $A^{a(f)}_k(\vec{x})$ are static classical gauge fields, and the state
$|A\rangle$ is such that
\begin{equation}
\label{stateA}
\langle A | \Psi \rangle = \Psi[A] \,.
\end{equation}
From Eq.~(\ref{Zeta}), inserting an orthonormal basis $|\Psi_n\rangle$ of
gauge invariant energy eigenstates, it follows:
\begin{equation}
\label{Zetaortho}
{\mathcal{Z}}[A^{(f)},A^{(i)}] = \sum_n e^{-E_n T} \Psi_n[A^{(f)}] 
\Psi^{*}[A^{(i)}] \,.
\end{equation}
Note that we are interested in the lattice version of the Schr\"odinger
functional, so that it makes sense to perform a discrete sum in 
Eq.~(\ref{Zetaortho}) for the spectrum is discrete in a finite volume.
Eq.~(\ref{Zetaortho}) shows that the Schr\"odinger functional is invariant
under arbitrary gauge transformations of the fields $A^{(f)}$ and 
$A^{(i)}$.

Using standard formal manipulations and the gauge invariance of the
Schr\"odinger functional it is easy to rewrite 
${\mathcal{Z}}[A^{(f)},A^{(i)}]$ as a functional integral~\cite{Rossi1980,Gross1981}
\begin{equation}
\label{Zetaint}
{\mathcal{Z}}[A^{(f)},A^{(i)}] = \int {\mathcal{D}}A \;
e^{-\int_0^T dx_4 \, \int d^3x \, {\mathcal{L}}_{YM}(x)}
\end{equation}
with the constraints:
\begin{eqnarray}
\label{constraints}
A_\mu(x_0=0) = & A^{(i)}_\mu \nonumber \\
\\
A_\mu(x_0=T) = & A^{(f)}_\mu \nonumber
\end{eqnarray}
Strictly speaking we should include in Eq.~(\ref{Zetaint}) the sum over
topological inequivalent classes. However, it turns out that~\cite{Luescher}
on the lattice such an average is not needed because the functional
integral Eq.~(\ref{Zetaint}) is already invariant under arbitrary
gauge transformations of $A_\mu^{(i)}$ and $A_\mu^{(f)}$.

On the lattice the natural relation between the continuum gauge fields and 
the corresponding lattice links is given by
\begin{equation}
\label{links}
U_\mu = {\mathrm P} \exp\left\{ 
iag \int_0^1 dt \, A_\mu(x+at\hat{\mu}) \right\}
\end{equation}
where ${\mathrm P}$ is the path-ordering operator, $a$ is the lattice
spacing and $g$ the gauge coupling constant.

The lattice implementation of the Schr\"odinger functional, Eq.~(\ref{Zetaint}),
is now straightforward:
\begin{equation}
\label{Zetalatt}
{\mathcal{Z}}[U^{(f)},U^{(i)}] = \int {\mathcal{D}}U \; e^{-S} \,.
\end{equation}
In Eq.~(\ref{Zetalatt}) the functional integration is done over
the links $U_\mu(x)$ with the fixed boundary values:
\begin{equation}
\label{boundary}
U(x)|_{x_4=0} = U^{(i)}\,, \quad U(x)|_{x_4=T} = U^{(f)} \,.
\end{equation}
Moreover the action $S$ is the standard Wilson action modified to take into 
account the boundaries at $x_4=0,T$~\cite{Luescher}:
\begin{equation}
S = \frac{1}{g^2} \sum_{x,\mu>\nu} W_{\mu\nu}(x) \, 
{\mathrm Tr} [1 - U_{\mu\nu}(x)]
\end{equation}
where $U_{\mu\nu}(x)$ are the plaquettes in the $(\mu,\nu)$-plane and
\begin{equation}
\label{Wmunu}
W_{\mu\nu}(x) = \left\{ 
\begin{array}{ll}
1/2  &  \mbox{spatial plaquettes at $x_4=0,T$} \\
1    &  \mbox{otherwise}
\end{array}
\right. \,.
\end{equation}
Moreover, it is possible to improve the lattice action $S$ by modifying 
the weights $W_{\mu\nu}$'s~\cite{Luescher}.
Note that, due to the fact that $U^{(i)} \ne U^{(f)}$, one cannot impose
periodic boundary conditions in the Euclidean time direction. On the other hand
one can assume periodic boundary conditions in the spatial directions. 

Let us consider, now, a static external background field 
$\vec{A}^{\mathrm{ext}}(\vec{x}) =  \vec{A}^{\mathrm{ext}}_a(\vec{x}) \lambda_a/2$,
where $\lambda_a/2$ are the generators of the SU(N) Lie algebra. We introduce, now,
a new functional:
\begin{equation}
\label{Gamma}
\Gamma[\vec{A}^{\mathrm{ext}}] = -\frac{1}{T} \ln \left\{ 
\frac{{\mathcal{Z}}[U^{\mathrm{ext}}]}{{\mathcal{Z}}(0)} \right\} \,,
\end{equation}
where
\begin{equation}
\label{ZetaUext}
{\mathcal{Z}}[U^{\mathrm{ext}}] = {\mathcal{Z}}[U^{\mathrm{ext}},U^{\mathrm{ext}}] \,,
\end{equation}
and ${\mathcal{Z}}[0]$ means the Schr\"odinger functional Eq.~(\ref{ZetaUext}) 
without external background field ($U^{\mathrm{ext}}_\mu=1$). The
lattice link $U^{\mathrm{ext}}$ is obtained from the continuum background
field $\vec{A}^{\mathrm{ext}}$ through Eq.~(\ref{links}).

From the previous discussion it is clear that $\Gamma[\vec{A}^{\mathrm{ext}}]$
is invariant for lattice gauge transformations of the external links 
$U^{\mathrm{ext}}_\mu$. Morever, from Eq.~(\ref{Zetaortho}) it follows
that
\begin{equation}
\label{limit}
\lim_{T \to \infty} \Gamma[\vec{A}^{\mathrm{ext}}] = 
E_0[\vec{A}^{\mathrm{ext}}] - E_0[0]
\end{equation}
where $E_0[\vec{A}^{\mathrm{ext}}]$ is the vacuuum energy in presence
of the external background field. In other words $\Gamma[\vec{A}^{\mathrm{ext}}]$
is the lattice gauge-invariant effective action for the static 
background field $\vec{A}^{\mathrm{ext}}$. In particular, if we consider
background fields that give rise to constant field strength, then due 
to the gauge invariance it is easy to show that $\Gamma[\vec{A}^{\mathrm{ext}}]$
is proportional to the spatial volume $V$. In this case one is interested in
the density of the effective action:
\begin{equation}
\label{density}
\varepsilon[\vec{A}^{\mathrm{ext}}] = -\frac{1}{\Omega} 
\ln \left[ \frac{{\mathcal{Z}}[A^{\mathrm{ext}}]}{{\mathcal{Z}}[0]} \right] \,,
\end{equation}
where $\Omega = V \cdot T$. We stress that our definition of the lattice effective
action uses the lattice Schr\"odinger functional with the same boundary fields at 
$x_4=0$ and $x_4=T$. So that we can glue the two hyperplanes $x_4=0$ and $x_4=T$
together. This way we end up in a lattice with periodic boundary conditions in
the time direction too. Therefore our lattice Schr\"odinger functional turns out
to be
\begin{equation}
\label{Zetal}
{\mathcal{Z}}[U^{\mathrm{ext}}] = \int {\mathcal{D}}U \; e^{-S} \;,
\end{equation}
where the functional integral is defined over a four-dimensional hypertorus 
with the ``cold-wall''
\begin{equation}
\label{coldwall}
U_\mu(x)|_{x_4=0} = U^{\mathrm{ext}}_\mu  \,.
\end{equation}
Moreover, due to the lacking of free boundaries, the lattice action in Eq.~(\ref{Zetal})
is now the familiar Wilson action
\begin{equation}
\label{SWilson}
S = S_W = \frac{1}{g^2} \sum_{x,\mu>\nu} Tr[1-U_{\mu\nu}(x)] \,.
\end{equation}
In this paper we study the properties of the gauge invariant lattice effective 
action in pure gauge non abelian theories. In particular we consider the
SU(2) gauge theory in presence of constant Abelian chromomagnetic field both
at zero and finite temperature. The plan of the paper is as follows. In Sect.~II
we consider the SU(2) gauge theory on the lattice in presence of constant Abelian
chromomagnetic background field.
Section~III is devoted to the discussion of the Nielsen-Olesen instability
on the lattice. In Sect.~IV we present the numerical results of the Monte Carlo
simulations at zero temperature~\cite{lat98}, while Sect.~V comprises the
finite temperature simulations. Finally our conclusions are drawn in Sect.~VI.

\renewcommand{\thesection}{\normalsize{\Roman{section}.}}
\section{\normalsize{SU(2) IN A CONSTANT ABELIAN CHROMOMAGNETIC FIELD}}
\renewcommand{\thesection}{\arabic{section}}

In this paper we are interested in the case of a constant Abelian chromomagnetic field.
Let us consider the SU(2) gauge theory. In the continuum we have:
\begin{equation}
\label{field}
\vec{A}^{\mathrm{ext}}_a(\vec{x}) =  \vec{A}^{\mathrm{ext}}(\vec{x}) \delta_{a,3} \,, \quad
\vec{A}^{\mathrm{ext}}_k(\vec{x}) =  \delta_{k,2} x_1 H \,.
\end{equation}
The external links corresponding to $\vec{A}^{\mathrm{ext}}_a(\vec{x})$ are easily evaluated
from Eq.~(\ref{links}):
\begin{eqnarray}
\label{extlinks}
U^{\mathrm{ext}}_1(\vec{x},0) = & U^{\mathrm{ext}}_3(\vec{x},0) = U^{\mathrm{ext}}_4(\vec{x},0)
= {\mathbf{1}} \nonumber \\
\\
U^{\mathrm{ext}}_2(\vec{x},0) = & \cos\left( \frac{agHx_1}{2} \right) + 
i \sigma^3 \sin\left( \frac{agHx_1}{2} \right)   \,. \nonumber
\end{eqnarray}
Our Schr\"odinger functional ${\mathcal{Z}}[\vec{A}^{\mathrm{ext}}]$ is defined on a lattice
with periodic boundary conditions, so that we impose that:
\begin{equation}
\label{pbcs}
U_2(x_1,x_2,x_3,x_4) = U_2(x_1+L_1,x_2,x_3,x_4) \,,
\end{equation}
where $L_1$ is the lattice extension in the $x_1$ direction (in lattice units).
As a consequence the magnetic field $H$ turns out to be quantized:
\begin{equation}
\label{quant}
\frac{a^2 g H}{2} = \frac{2 \pi}{L_1} n_{\mathrm{ext}} \,,
\end{equation}
with $n_{\mathrm{ext}}$ integer.

According to our previous discussion in evaluating the lattice functional integral
Eq.~(\ref{Zetal}) we impose that the links belonging to the time slice $x_4=0$ 
are frozen to the configuration Eq.~(\ref{extlinks}). Moreover we impose also that 
the links at the spatial boundaries are fixed according to Eq.~(\ref{extlinks}).
In the continuum this last condition amounts to the usual requirement that the fluctuations
over the background fields vanish at  infinity. An alternative possibility is given
by constraining the links belonging to the the time slice $x_4=0$ and those at
the spatial boundaries to the condition
\begin{equation}
\label{newpbcs}
U_2(x) = U_2^{\mathrm{ext}}(\vec{x}) \,,
\end{equation}
while the links $U_\mu(x)$ with $\mu \ne 1,2$ are unconstrained.
The main advantage of the condition~(\ref{newpbcs}) resides in the fact that 
the time-like plaquettes nearest the frozen hypersurface $x_4=0$ behave symmetrically
in the update procedure. Obviously in the thermodynamic limit both conditions 
should agree as the effective action is concerned.

\renewcommand{\thesection}{\normalsize{\Roman{section}.}}
\section{\normalsize{THE NIELSEN-OLESEN UNSTABILITY ON THE LATTICE}}
\renewcommand{\thesection}{\arabic{section}}

As it is well known in the continuum the perturbative evaluation of the effective
action for the constant Abelian chromomagnetic field faces with the problem of 
Nielsen-Olesen unstable modes~\cite{Nielsen1978}. Let us briefly discuss the origin
of the unstable modes in the continuum. 

In order to evaluate the effective action in the continuum one writes
\begin{equation}
\label{Apiueta}
A^a_\mu(x) = \overline{A}^a_\mu(x) + \eta^a_\mu(x) \,,
\end{equation}
where $\overline{A}^a_\mu(x) = \delta_{\mu 2} \delta^{a 3} x_1 H$ and $\eta^a_\mu(x)$
is the quantum fluctuation over the background field. In the background gauge
\begin{equation}
\label{bgauge}
[\delta^{ab} \partial_\mu - g^{abc} \overline{A}^c_\mu(x)] \eta^b_\mu(x) = 0
\end{equation}
we rewrite the pure gauge action in the one-loop approximation as:
\begin{equation}
\label{Soneloop}
S = S_{\mathrm{class}} + \frac{1}{2} \int d^4x \, \eta^a_\mu(x) \, {\mathcal{O}}^{ab}_{\mu\nu}(x) \,.
\end{equation}
The one-loop effective action can be obtained by performing the Gaussian integration over the
quantum fluctuations and including the Faddeev-Popov determinant. However, if we solve
the eigenvalue equation
\begin{equation}
\label{eigenvalue}
{\mathcal{O}}^{ab}_{\mu\nu} \phi^b_\nu(x) = \lambda  \phi^a_\mu(x)
\end{equation}
then we find that there are negative eigenvalues:
\begin{equation}
\label{negative}
\lambda_u = p^2_0 + p^2_3 - gH  \,.
\end{equation}
As a matter of fact $\lambda_u<0$ when $gH> p^2_0 + p^2_3$. If we perform formally
the Gaussian functional integration in the one-loop approximation then the effective 
action picks up an imaginary part. The point is that in the functional integration 
over the unstable modes one must include the positive quartic term. It turns out that 
the unstable modes behave like a two-dimensional tachyonic charged scalar field.
Thus the dynamics of the unstable modes resemble the dynamical Higgs mechanism. As
a consequence the response of the gauge system to the external field
turns out to be strong even in the nominally perturbative regime~\cite{Cea1988}.

In order to ascertain if the Nielsen-Olesen one-loop instability survives
the lattice regularization one should evaluate the Schr\"odinger functional 
Eq.~(\ref{ZetaUext}) in the weak coupling region. To this end we write the lattice
version of Eq.~(\ref{Apiueta}):
\begin{equation}
\label{Umu}
U_\mu(x) = \exp( i\, a\, g\, q_\mu(x) ) \,\,U_\mu^{\mathrm{ext}}  \,,
\end{equation}
where the fluctuations $q_\mu(x) = q^a_\mu(x) \sigma^a/2$ satisfy the boundary
condition
\begin{equation}
\label{qmu}
q_\mu(x)|_{x_4=0} = 0 \,.
\end{equation}
Inserting Eq.~(\ref{Umu}) into the plaquette
\begin{equation}
\label{Umunu}
U_{\mu\nu}(x) = U_\mu(x) U_\nu(x+\hat{\mu}a) U^\dagger_\mu(x +\hat{\nu}a)
U^\dagger_\nu(x) \,,
\end{equation}
we rewrite the Wilson action 
\begin{equation}
\label{wilsact}
S_W = \frac{4}{g^2} \sum_x \sum_{\mu>\nu} \left[ 1 - \frac{1}{2} {\mathrm{tr}}
U_{\mu\nu}(x) \right]
\end{equation}
in the quadratic approximation as:
\begin{equation}
\label{Squadratic}
S_W = S^{\mathrm{ext}} + S^{(2)}
\end{equation}
where
\begin{equation}
\label{Sext}
S^{\mathrm{ext}} = \frac{4 \Omega}{g^2} \left[ 1 - \cos(\frac{gHa^2}{2}) \right] \,,
\end{equation}
with $\Omega = L_1 \times  L_2 \times    L_3 \times  L_4 $ the lattice volume. Note that 
the external action $S^{\mathrm{ext}}$ in the naive continuum limit reduces to
the classical action
\begin{equation}
\label{Sclass}
S^{\mathrm{cl}} = V T \frac{H^2}{2} \,
\end{equation}
As concerns the quadratic action $S^{(2)}$, following the method of Ref.~\cite{Dashen1981}
a standard calculation gives
%
\begin{eqnarray}
\label{S2}
S^{(2)} = && a^4 \sum_{x, \mu>\nu} {\mathrm{Tr}} \left\{ [D_\mu q_\nu(x) -
D_\nu q_\mu(x) ]^2 U^{\mathrm{ext}}_{\mu\nu}(x) \right\} \nonumber \\
&& - a^4  \sum_{x, \mu>\nu} {\mathrm{Tr}} \left\{ [D_\nu q_\mu(x),
D_\mu q_\nu(x) ] U^{\mathrm{ext}}_{\mu\nu}(x) \right\} \nonumber \\
&& -2 a^2 \sum_{x, \mu>\nu} {\mathrm{Tr}} \left\{ [q_\mu(x),q_\nu(x) ]
 U^{\mathrm{ext}}_{\mu\nu}(x) \right\} \nonumber \\
&& -a^3 \sum_{x, \mu>\nu} {\mathrm{Tr}} \left\{ \left( [D_\nu q_\mu(x), q_\mu(x)] 
\right. \right. \nonumber \\
&& \left. \left. \qquad \qquad \qquad 
-[D_\mu q_\nu(x), q_\nu(x)] \right) U^{\mathrm{ext}}_{\mu\nu}(x) \right\} \,
\end{eqnarray}
where $D_\mu$ is the lattice covariant derivative in the external background 
$U^{\mathrm{ext}}_\mu(x)$:
\begin{equation}
\label{Dmu}
D_\mu f(x) = \frac{1}{a} \left[ U^{\mathrm{ext}}_\mu(x) f(x + \hat{\mu}a) 
U^{\mathrm{ext}\dagger}_\mu(x) -f(x) \right] \,.
\end{equation}
Observing that 
\begin{equation}
\label{Umunuext}
U^{\mathrm{ext}}_{\mu\nu} = G_{\mu\nu} + i H_{\mu\nu}
\end{equation}
with
\begin{equation}
\label{Gmunu}
G_{\mu\nu} = \left\{ 
\begin{array}{ll}
{\mathbf{1}}              &  (\mu,\nu) \ne (1,2) \\
\cos(\frac{a^2 g B}{2})     &  (\mu,\nu) = (1,2)
\end{array}
\right. \,,
\end{equation}
\begin{equation}
\label{Hmunu}
H_{\mu\nu} = \left\{ 
\begin{array}{ll}
0              &  (\mu,\nu) \ne (1,2) \\
\sin(\frac{a^2 g B}{2})     &  (\mu,\nu) = (1,2)
\end{array}
\right. \,,
\end{equation}
and integrating by parts, we rewrite the quadratic action as:
%
\begin{eqnarray}
\label{S2int}
S^{(2)} = && 2 a^4 \sum_{x, \mu<\nu} G_{\mu\nu} {\mathrm{Tr}} 
[q_\mu(x) (D^*_\nu D_\mu - \delta_{\mu\nu} D^*_\sigma D_\sigma) q_\nu(x)]
\nonumber \\
&& + 2i a^4  \sum_{x, \mu>\nu} {\mathrm{Tr}} [q_\mu(x) D^*_\nu D_\mu q_\nu(x) 
H_{\mu\nu}(x) ]
\nonumber \\
&& +2 i a^3 \sum_{x, \mu>\nu} {\mathrm{Tr}} 
[q_\mu(x) (D^*_\nu+D_\nu) q_\mu(x) H_{\mu\nu}(x) ]
\nonumber \\
&& -4 i a^2 \sum_{x, \mu>\nu} {\mathrm{Tr}} 
[q_\nu(x) q_\mu(x) H_{\mu\nu}(x) ]
\,,
\end{eqnarray}
where
\begin{eqnarray}
\label{Dstarnu}
D^*_\nu f(x) = && \frac{1}{a}  \left[ f(x) - \left( U^{\mathrm{ext}}_\mu \right)^{-1}(x
- a \hat{\mu})  \right. \nonumber \\
&& \qquad \times \left.  f(x-a\hat{\mu}) U^{\mathrm{ext}}_\mu(x-a\hat{\mu}) \right] \,.
\end{eqnarray}
Taking into account that 
\begin{equation}
\label{qmua}
q_\mu(x) = q_\mu^a(x) \frac{\sigma^a}{2}
\end{equation}
and using Eq.~(\ref{extlinks}) we perform the trace over the color indexes. After a rather
long but otherwise elementary calculation we get
\begin{equation}
\label{S2q3}
S^{(2)} = S^{(2)}(q^3) + S^{(2)}(q^+,q^-) \,,
\end{equation}
where
\begin{equation}
\label{qmupm}
q^\pm_\mu(x) = q^1_\mu(x) \pm q^2_\mu(x) \,.
\end{equation}
We have
\begin{eqnarray}
\label{S3q2}
S^{(2)}(q^3) = && \frac{a^4}{2} \sum_x \sum_{\mu\nu} G_{\mu\nu} [ q^3_\mu(x) \Delta^*_\nu
\Delta_\mu q^3_\nu(x) \nonumber \\
&& \qquad \qquad \qquad - q^3_\nu(x) \Delta^*_\mu
\Delta_\mu q^3_\nu(x) ] \,,
\end{eqnarray}
where
\begin{eqnarray}
\label{Deltamu}
\Delta_\mu f(x) &=& \frac{1}{a} [f(x+\hat{\mu}a) - f(x) ]  \,, \nonumber \\
 \\
\Delta^*_\mu f(x) &=& \frac{1}{a} [f(x) - f(x-\hat{\mu}a)]  \nonumber \,.
\end{eqnarray}
Moreover we have:
%
\begin{eqnarray}
\label{S2qpiuqmeno}
& &  S^{(2)}(q^+,q^-) = 
\frac{a^4}{4} \sum_x \sum_{\mu\nu} G_{\mu\nu} \left( 
q_\mu^-(x) \overline{{\mathcal{D}}}^+_\nu {\mathcal{D}}^+_\mu q_\nu^+(x) \right. \nonumber \\ 
& & \left. +  q_\mu^+(x) \overline{{\mathcal{D}}}^-_\nu {\mathcal{D}}^-_\mu q_\nu^-(x) 
- q_\nu^-(x) \overline{{\mathcal{D}}}^+_\mu {\mathcal{D}}^+_\mu q_\nu^+(x)  \right. \nonumber \\
& & \left. - q_\nu^+(x) \overline{{\mathcal{D}}}^-_\mu {\mathcal{D}}^-_\mu q_\nu^-(x) \right) \nonumber \\
& & + \frac{ia^4}{2} \sum_x 
\left( q_1^-(x) \overline{{\mathcal{D}}}^+_2 {\mathcal{D}}^+_1 q_2^+(x)
- q_1^+(x) \overline{{\mathcal{D}}}^-_2 \overline{{\mathcal{D}}}^-_1 q_2^-(x) \right) H_{12} \nonumber \\
& & +  \frac{i a^3}{2} \sum_x \left( 
q^-_1(x)  \overline{{\mathcal{D}}}^+_2 {\mathcal{D}}^+_1 q_2^+(x)
- q_1^+(x) \overline{{\mathcal{D}}}^-_2 {\mathcal{D}}^-_1 q_2^-(x) \right) H_{12} \nonumber \\
& & + ia^2 \sum_x \left( q_2^-(x) q_1^+(x) - q_1^-(x) q_2^+(x) \right) H_{12} \,,
\end{eqnarray}
with
%
\begin{eqnarray}
\label{Dmupiumeno}
{\mathcal{D}}_\mu^\pm f(x) & = & \Delta_\mu f(x) \pm \frac{2i}{a} \delta_{\mu 2}
\sin \left( \frac{a g H}{2} x_1 \right) f(x + \hat{\mu}a) \,, \nonumber \\
 \\
\overline{{\mathcal{D}}}_\mu^\pm f(x) & = & \Delta^*_\mu f(x) \pm \frac{2i}{a} \delta_{\mu 2}
\sin \left( \frac{a g H}{2} x_1 \right) f(x - \hat{\mu}a) \,. \nonumber
\end{eqnarray}
Obviously we need, now, to fix the gauge. To this end we add a gauge fixing term
to the action and the associated Faddeev-Popov ghost field action.
We use the background gauge condition:
\begin{equation}
\label{backgroundgauge}
\sum_\mu D_\mu q_\mu(x) = 0 \,.
\end{equation}
In the Landau gauge the gauge-fixing term in the one-loop approximation
is given by
\begin{equation}
\label{S2gf}
S^{(2)}_{gf} = a^4 \sum_x {\mathrm{Tr}} \left[\sum_\mu D_\mu q_\mu(x) \right]^2 \,.
\end{equation}
Moreover, in the same approximation we get the following Faddeev-Popov contribution
\begin{equation}
\label{S2FP}
S^{(2)}_{F-P} = -\ln {\mathrm{Det}}\left[ -\sum_\mu D_\mu^* D_\mu \right] \,.
\end{equation}
The lattice version of the continuum operator ${\mathcal{O}}^{ab}_{\mu\nu}$ in
Eq.~(\ref{Soneloop}) can be extracted from Eqs.~(\ref{qmupm}), (\ref{S2qpiuqmeno}), 
and (\ref{S2gf}). 
Unlike the continuum case it is not possible to solve in closed form the lattice
version of the eigenvalues equations Eq.~(\ref{eigenvalue}). However, if we neglect
the irrelevant terms and keep only the contributions that survive in the naive 
continuum limit $a \to 0$, then we were able to solve the eigenvalue equations
and obtain the spectrum. In this approximation we replace $G_{\mu\nu}$ with 
the identity. So that $S^{(2)}(q^3)$ 
does not depend on the background field and
we can discard it. Moreover the sum of $S^{(2)}(q^+,q^-)$ 
and $S^{(2)}_{gf}$
simplifies considerably. We get
%
\begin{eqnarray}
\label{SqpqmpiuSgf}
& & S^{(2)}(q^+,q^-) + S^{(2)}_{gf} = \frac{a^4}{2}
\sum_x \sum_\mu q^+_\mu(x)[-{\mathcal{O}}_1] q^-_\mu(x) + \mathrm{h.c.} \nonumber \\
& & + \frac{a^4}{2} \sum_x \sum_{\mu,\nu} (\delta_{\mu1} \delta_{\nu2} -
\delta_{\mu2} \delta_{\nu1}) q^-_\mu(x)  {\mathcal{O}}_2 q^+_\nu(x) + \mathrm{h.c.} \,,
\end{eqnarray}
where we restricted the quantum fluctuations to the class of function
\begin{equation}
\label{class}
q(x_1,x_2,x_3,x_4) = \sin(p_4x_4) e^{i(p_2x_2+p_3x_3)} f(x_1) \,.
\end{equation}
Note that the class of functions Eq.~(\ref{class}) is relevant for the constraint
Eq.~(\ref{extlinks}). Similar results can be obtained with the constraints 
Eq.~(\ref{newpbcs}).
The periodic boundary conditions imply that
\begin{equation}
\label{fx1}
f(x_1+L_1)=f(x_1) 
\end{equation}
\begin{equation}
\label{pmu}
p_\mu = \frac{2 \pi}{L_\mu} n_\mu  \qquad \mu=2,3,4 \,,
\end{equation}
and $n_\mu$ integer.
Within the class of functions Eq.~(\ref{class}) we have
\begin{eqnarray}
\label{theta1}
\lefteqn{-{\mathcal{O}}_1  = - \Delta_1^* \Delta_1 
+ \frac{4}{a^2} \sin^2 \left( \frac{a g H x_1}{2} \right)}
\nonumber \\
& & + \frac{2}{a^2} \sum_{\mu=1}^4 (1 -\cos p_\mu a ) 
- \frac{2}{a^2} \sin(a g H x_1) \sin(p_2 a) \nonumber \\
& &  - \frac{4}{a^2} 
\sin^2 \left( \frac{a g H x_1}{2} \right) ( 1 - \cos(p_2 a)) \,,
\end{eqnarray} 
\begin{eqnarray}
\label{theta2}
\lefteqn{{\mathcal{O}}_2  =  \frac{2 i}{a^2} \sin( \frac{a^2 g H}{2} ) + \frac{i}{a^2} 
\sin(g H a^2) \cos(g H a x_1)}\nonumber\\
& & \qquad - \frac{2}{a^2} \sin(a g H x_1) \sin(p_2 a)  \frac{\Delta_1 + \Delta_1^*}{2} \,.
\end{eqnarray} 
By keeping only the relevant terms, the operators ${\mathcal{O}}_1$ and ${\mathcal{O}}_2$ further
simplify as:
\begin{eqnarray}
\label{theta1simp}
-{\mathcal{O}}_1 & = & - \Delta_1^* \Delta_1 
+ \frac{4}{a^2} \sin^2 \left( \frac{a g H x_1}{2} \right) 
-\frac{2}{a^2} \sin(a g H x_1) \sin(p_2 a) \nonumber\\
& & + \frac{2}{a^2} \sum_{\mu=1}^4 (1 - \cos( p_2 a)) \,,
\end{eqnarray}
\begin{eqnarray}
\label{theta2simp}
{\mathcal{O}}_2 & = & \frac{2 i}{a^2} \sin( \frac{a^2 g H}{2} ) + \frac{i}{a^2} 
\sin(g H a^2)  \nonumber \\
& & \simeq \frac{2 i}{a^2} \sin( \frac{a^2 g H}{2} ) \,.
\end{eqnarray}
Introducing the complex scalar fields 
\begin{eqnarray}
\label{cscalfield}
\phi = q^-_+(x) \quad, \quad \phi^* = q^+_-(x)  \nonumber \\
 \\
\psi = q^-_-(x) \quad, \quad \psi^* = q^+_+(x)  \nonumber
\end{eqnarray}
where $q_\pm = \frac{1}{\sqrt{2}} (q_1 \pm q_2)$, we get 
\begin{eqnarray}
\label{s2qpiuqmeno}
S^{(2)}&(&q^+,q^-) + S^{(2)}_{g-f}  = \frac{a^4}{2} \sum_x \sum_{\nu=3,4} q^+_\nu(x)[-{\mathcal{O}}_1] q^-_\nu 
+  \, {\mathrm{h.c.}} \nonumber \\
& & +  \frac{a^4}{2} \sum_x \psi^*(x) [-{\mathcal{O}}_1 + {\mathcal{O}}_2] \psi(x) 
+ \, {\mathrm{h.c.}} \nonumber \\
& & +  \frac{a^4}{2} \sum_x \phi^*(x) [-{\mathcal{O}}_1 -{\mathcal{O}}_2] \phi(x) 
+  \, {\mathrm{h.c.}}
\end{eqnarray}
It is easy to verify that the contribution to the one-loop effective action 
due to the fluctuating fields $q_\mu^\pm, \mu=3,4$ cancels the one due to the 
Faddeev-Popov determinant. So that we are left with the following
quadratic action:
\begin{eqnarray}
\label{quadract}
S^{(2)}(\psi,\phi) &=& a^4 \sum_x \psi^*(x) [-{\mathcal{O}}_1(x) + m^2] \psi(x)
\nonumber \\
& & + a^4 \sum_x \phi^*(x) [-{\mathcal{O}}_1(x) - m^2] \phi(x)
\end{eqnarray}
where
\begin{equation}
\label{mquadro}
m^2 = \frac{2}{a^2} \sin(a^2 g H) \,.
\end{equation}
Let us introduce the operators 
\begin{eqnarray}
\label{C-operators}
C & = & \Delta_1 - i \left[ \frac{ e^{i p_2 a} -1}{a} \right] + 
\frac{2}{a} \sin( \frac{gHax_1}{2} )  \\
C^* & = & -\Delta_1^* + i \left[ \frac{ e^{-i p_2 a} -1}{a} \right] + 
\frac{2}{a} \sin( \frac{gHax_1}{2} ) \,.
\end{eqnarray}
By keeping the leading terms in the continuum limit it is not too
hard to see that
\begin{equation}
\label{menotheta1}
-{\mathcal{O}}_1 = \frac{1}{2} [C^* C + C C^*] + \sum_{\nu=3}^4 \frac{2}{a^2} (1 - \cos p_\mu a)
\end{equation}
and
\begin{equation}
\label{ccstar}
[C,C^*] = \frac{4}{a^2} \sin( \frac{g H a^2}{2}) \,.
\end{equation}
So that we find that the eigenvalue equation 
\begin{equation}
\label{eigenveq}
-{\mathcal{O}}_1 f_\lambda(x_1) = \lambda f_\lambda(x_1)
\end{equation}
admits the solutions:
\begin{eqnarray}
\label{lambdau}
\lambda_n  & = &   \frac{4 n}{a^4} \sin(\frac{g H a^2}{2}) \nonumber \\
& & + \frac{2}{a^2} \sum_{\mu=3}^4 (1 - \cos p_\mu a) \,, n=0,1,2,\dots \,.
\end{eqnarray}
Note that the eigenvalues are degenerate. As a matter of fact the order of the degeneracy
of the Landau levels turns out to be
\begin{equation}
\label{degeneracy}
g = (\frac{L_1 a}{2 \pi}) (\frac{L_2 a}{2 \pi}) \frac{2}{a^2} \sin( \frac{gH a^2}{2}) \,.
\end{equation}
We have numerically checked that the approximate spectrum Eq.~(\ref{lambdau}) 
agrees quite well with the exact one as long as $L_1 \ge 32$ and for weak magnetic field.

From Eqs.~(\ref{eigenveq}) and~(\ref{lambdau}) we find the following eigenvalues:
\begin{eqnarray}
\label{eigenvalues}
\lambda_n^\psi & = & \sum_{\nu=3}^4 \frac{2}{a^2} (1 -\cos p_\nu a) 
\nonumber \\ 
& & + (2 n + 3) \frac{2}{a^2} \sin^2 \frac{gH a^2}{2} \,, n=0,1,2,\dots   \,,\\
\lambda_n^\phi & = & \sum_{\nu=3}^4 \frac{2}{a^2} (1 -\cos p_\nu a) 
\nonumber \\
& & + (2 n - 1) \frac{2}{a^2} \sin^2 \frac{gH a^2}{2} \,, n=0,1,2,\dots \,.
\end{eqnarray}
It is now evident that the $\phi$-mode with $n=0$ is the Nielsen-Olesen mode with eigenvalues
\begin{eqnarray}
\label{Nielsen-Olesen-mode}
\lambda_u & = & \frac{2}{a^2} (1 - \cos p_3 a) +  \frac{2}{a^2} (1 - \cos p_4 a) 
\nonumber \\  & & -  \frac{2}{a^2} \sin^2 (\frac{g H a^2}{2} )
\end{eqnarray}
which is the discretized version of Eq.~(\ref{negative}).

\renewcommand{\thesection}{\normalsize{\Roman{section}.}}
\section{\normalsize{MONTE CARLO SIMULATIONS: $T=0$}}
\renewcommand{\thesection}{\arabic{section}}

Our numerical simulations have been done on a lattice of size $L_1 L_2 L_3 L_4$ with
periodic boundary conditions. In order to project onto the ground state according
to Eq.(\ref{limit}) we need $L_4 \gg 1$. Moreover in order to be close to the continuum
limit Eqs.(\ref{Sext}) and~(\ref{quant}) imply also $L_1 \gg 1$. As a consequence
we performed the numerical simulations on lattices with $L_1 = L_4 = 32$.
The transverse size of the lattice $L_\perp = L_2 = L_3$ has been varied from $L_\perp=6$
up to  $L_\perp=32$.
We are interested in the density of the effective action Eq.~(\ref{density}).
We face with the problem of computing a partition function which is the 
exponential of an extensive quantity~\cite{Hasenfratz1990}. To avoid this problem
we consider the derivative of $\varepsilon[\vec{A}^{\mathrm{ext}}]$ with
respect to $\beta$ by taking $n_{\mathrm{ext}}$ (i.e. $gH$) fixed (see Eq.~(\ref{quant})).
From Eqs.(\ref{density}), (\ref{Zetal}), and (\ref{SWilson}) it follows:
\begin{eqnarray}
\label{epsilonprimo}
\lefteqn{
\varepsilon^\prime[\vec{A}^{\mathrm{ext}}]  =  
\frac{\partial \varepsilon[\vec{A}^{\mathrm{ext}}]}{\partial \beta} 
 = -\frac{1}{\Omega} \left[ \frac{1}{{\mathcal{Z}}[U^{\mathrm{ext}}]} 
\frac{\partial {\mathcal{Z}}[U^{\mathrm{ext}}]}{\partial \beta} \right. 
} 
\nonumber \\
& & \left. - \frac{1}{{\mathcal{Z}}[0]} 
\frac{\partial {\mathcal{Z}}[0]}{\partial \beta} \right]  
= \left\langle \frac{1}{\Omega} \sum_{x, \mu> \nu} \frac{1}{2} {\mathrm{Tr}} 
U_{\mu \nu}(x) \right\rangle_0 \nonumber \\
& & - \left\langle \frac{1}{\Omega} \sum_{x, \mu> \nu} \frac{1}{2} {\mathrm{Tr}} 
U_{\mu \nu}(x) \right\rangle_{\vec{A}^{\mathrm{ext}}}\,,
\end{eqnarray}
where the subscripts on the average indicate the value of the external links
at the boundaries. Obviously $\varepsilon[\vec{A}^{\mathrm{ext}}]$ can be obtained
by a numerical integration in $\beta$ 
\begin{equation}
\label{numericalint}
\varepsilon[\vec{A}^{\mathrm{ext}}, \beta] = 
\int_0^\beta \, d\beta^{\prime} \, 
\varepsilon^{\prime}[\vec{A}^{\mathrm{ext}}, \beta^{\prime}]   \,,
\end{equation}
where we have taken into account that 
$\lim_{\beta \to 0} \varepsilon[\vec{A}^{\mathrm{ext}}, \beta] = 0$.

It is evident that the contributions to $\varepsilon^{\prime}[\vec{A}^{\mathrm{ext}}]$
due to the frozen time-slice at $x_4=0$ and to the fixed links at the spatial 
boundaries must be subtracted. In other words, only the dynamical links must
be taken into account in evaluating $\varepsilon^{\prime}[\vec{A}^{\mathrm{ext}}]$.
We recall that $\Omega=L_1 L_2 L_3 L_4$ is the total number of lattice sites
(i.e. the lattice volume) belonging to the lattice $\Lambda$. If we denote with
$\Omega_{\mathrm{ext}}$ the lattice sites whose links are fixed according to 
Eq.~(\ref{extlinks}):
\begin{eqnarray}
\label{omegaext}
\lefteqn{\Omega_{\mathrm{ext}}  = L_1 L_2 L_3} \nonumber \\
& & + (L_4-1) (L_1 L_2 L_3 -(L_1-2)(L_2-2)(L_3-2)) \,,
\end{eqnarray}
then the volume occupied by the ``internal'' lattice sites is given by
\begin{equation}
\label{omegaint}
\Omega_{\mathrm{int}} = \Omega - \Omega_{\mathrm{ext}} \,.
\end{equation}
Accordingly, we define the derivative of the internal energy density 
$\varepsilon^{\prime}_{\mathrm{int}}[\vec{A}^{\mathrm{ext}}]$ as 
\begin{eqnarray}
\label{deriv}
\lefteqn{
\varepsilon^{\prime}_{\mathrm{int}}[\vec{A}^{\mathrm{ext}}] =
\left \langle \frac{1}{\Omega_{\mathrm{int}}} 
\sum_{x \in \tilde{\Lambda},\mu > \lambda}
\frac{1}{2} {\mathrm{Tr}} U_{\mu\nu}(x) \right\rangle_0} \nonumber \\
& & -
\left\langle \frac{1}{\Omega_{\mathrm{int}}} 
\sum_{x \in \tilde{\Lambda},\mu > \lambda}
\frac{1}{2} {\mathrm{Tr}} U_{\mu\nu}(x) \right\rangle_{\vec{A}^{\mathrm{ext}}} \,,
\end{eqnarray}
where $\tilde{\Lambda}$ is the ensemble of the internal lattice sites.

We use the over-relaxed heat-bath algorithm to update the gauge configurations. 
Simulations have been performed by means of the APE100/Quadrics computer.
Since we are measuring a local quantity such as the plaquette, a low statistics
(from 1000 up to 5000 configurations) is required in order to get a good estimation 
of $\varepsilon^{\prime}_{\mathrm{int}}$.

In Figure 1 we display the derivative of the energy density normalized to the 
derivative of the external energy density:
\begin{equation}
\label{epsprimeext}
\varepsilon^{\prime}_{\mathrm{ext}} = 1 - \cos( \frac{g H}{2} ) = 
1 - \cos( \frac{2 \pi}{L_1} n_{\mathrm{ext}}) 
\end{equation}
versus $\beta$ for $L_1=L_4=32$ and $6 \le L_\perp \le 32$.
From Figure 1 we see that in the strong coupling region $\beta \lesssim 1$ 
the external background field is completely shielded. Moreover 
$\varepsilon^{\prime}_{\mathrm{int}}$ display a peak at $\beta \simeq 2.2$
resembling the behavior of the specific heat~\cite{Lautrup80,Engels97}. This is not surprising
since our previous studies~\cite{fsu1} in U(1) showed that  
$\varepsilon^{\prime}_{\mathrm{int}}$ behaves like a specific heat.

In the weak coupling region $\beta \gtrsim 3$ Fig.~1 shows that the ratio
$\varepsilon^{\prime}_{\mathrm{int}}/\varepsilon^{\prime}_{\mathrm{ext}}$ stays 
constant. Actually the constant does depend on $L_\perp$ and $n_{\mathrm{ext}}$.
Indeed in Fig.~1  the dependence on $L_\perp$ for fixed external magnetic field 
is evident. On the other hand in Fig.~2 we keep $L_\perp=32$ fixed and vary $n_{\mathrm{ext}}$.
We see clearly that the weak coupling plateau decreases by increasing the external field.
In order to extract $\varepsilon_{\mathrm{int}}(\beta,n_{\mathrm{ext}})$ we can
numerically integrate the data for 
$\varepsilon^{\prime}_{\mathrm{int}}(\beta,n_{\mathrm{ext}})
/\varepsilon^{\prime}_{\mathrm{ext}}$ using the trapezoidal rule
\begin{equation}
\label{trapez}
\varepsilon_{\mathrm{int}}(\beta,n_{\mathrm{ext}})  = 
\varepsilon^{\prime}_{\mathrm{ext}} \,
\int_0^\beta 
\frac{\varepsilon^{\prime}_{\mathrm{int}}(\beta,n_{\mathrm{ext}})}{\varepsilon^{\prime}_{\mathrm{ext}}}
\,d\beta^{\prime}   \,.
\end{equation}
In Figure 3 we display 
$\varepsilon_{\mathrm{int}}(\beta,n_{\mathrm{ext}})$ 
obtained from Eq.~(\ref{trapez}).
The plateau of the derivative of the internal energy density in the weak coupling region results in
a linear rising term in the energy density.
For $\beta \gg 1$ we get 
\begin{equation}
\label{betagg1}
\varepsilon_{\mathrm{int}}(\beta,n_{\mathrm{ext}}) \simeq \beta \varepsilon^{\prime}_{\mathrm{ext}} 
a(n_{\mathrm{ext}}) \,.
\end{equation}
Moreover for $\beta \gg 1$ we get also
\begin{equation}
\label{classical}
\beta \, \varepsilon^{\prime}_{\mathrm{int}} = \beta \,(1 -\cos \frac{2 \pi}{L_1} n_{\mathrm{ext}}) 
\simeq \frac{1}{2} H^2 \,.
\end{equation}
So that in the weak coupling region
\begin{equation}
\label{weakcoupling}
\varepsilon_{\mathrm{int}}(\beta,n_{\mathrm{ext}}) \simeq a(n_{\mathrm{ext}}) \frac{1}{2} H^2 \,,
\quad \beta \gg 1 \,.
\end{equation}
Figure 1 shows that $a(n_{\mathrm{ext}}) \simeq 1$ for $L_\perp \simeq 6-8$ and
$n_{\mathrm{ext}} = 1$. On the other hand  $a(n_{\mathrm{ext}})$ decreases by increasing 
$L_\perp$ or the external background field. This peculiar behavior can be compared with the 
Abelian case where we found that $a(n_{\mathrm{ext}}) \simeq 1$ independently on $L_\perp$
and $n_{\mathrm{ext}}$~\cite{metodo,fsu1}. Previous theoretical studies~\cite{Cea1988} suggested
that due to the presence of the Nielsen-Olesen modes the gauge system reacts strongly to
the external perturbation even in the nominally perturbative regime. It turns out
that the Nielsen-Olesen modes behave like a (1+1)-dimensional tachionic charged scalar field.
The condensation of these modes takes place only in the thermodynamic limit. As a consequence
the applied external background magnetic field is almost completely screened and there
is a dramatic reduction of the vacuum magnetic energy. Indeed it turns out that 
in the infinite volume limit the perturbative vacuum and the magnetic condensate vacuum are degenerate
for vanishing gauge coupling. 

On the lattice the Nielsen-Olesen modes display the one-loop instability when $\lambda_u$ 
given by Eq.~(\ref{Nielsen-Olesen-mode}) becomes negative. In the approximation 
adopted in Sect.~III we find that $\lambda_u$ gets negative by increasing
$L_\perp$ for fixed external field. Thus we can switch on and off the one-loop instability
by varying $L_\perp$. This has been also noticed by the Authors of Ref.~\cite{Levi-Polonyi}. 
For instance, by using Eqs.~(\ref{quant}), (\ref{pmu}),
and (\ref{Nielsen-Olesen-mode}) with $L_1=L_4=32$ and $n_{\mathrm{ext}}=1$ we find that 
$\lambda_u \lesssim 0$ for $L_\perp \gtrsim 11$.

Our numerical results in Fig.~1 show that for $n_{\mathrm{ext}}=1$ and $L_\perp \lesssim 10$
there is no the Nielsen-Olesen instability and the gauge system responds weakly to
the external perturbation in the weak coupling region. On the other hand, by increasing 
$L_\perp$ (see Fig.~1) or $n_{\mathrm{ext}}$ (see Fig.~2) we increment the lattice Nielsen-Olesen
modes. As a consequence we find a clear reduction of the vacuum energy density for both the 
peak values of $\varepsilon^{\prime}_{\mathrm{int}}(\beta,n_{\mathrm{ext}})$ and the
coefficient $a(n_{\mathrm{ext}})$ in Eq.~(\ref{weakcoupling}) decreases towards zero in the
thermodynamic limit. To see this we need to perform the infinite volume extrapolation.
We can extract more information from our numerical data by expressing them versus
\begin{equation}
\label{x-scaling}
x = \frac{a_H}{L_{\mathrm{eff}}} \,,
\end{equation}
where
\begin{equation}
\label{mag-length}
a_H = \sqrt{ \frac{2 \pi}{g H} } = \sqrt{ \frac{L_1}{2 n_{\mathrm{ext}}} }
\end{equation}
is the magnetic length and
\begin{equation}
\label{L-eff}
L_{\mathrm{eff}} = \Omega_{\mathrm{int}}^{1/4}
\end{equation}
is the lattice effective linear size.
Indeed we find that the data for 
$\varepsilon^{\prime}_{\mathrm{int}}(\beta,n_{\mathrm{ext}})/\varepsilon^{\prime}_{\mathrm{ext}}$
at the perturbative tail and at the peak for various lattice sizes and values of
$n_{\mathrm{ext}}$ can be expressed as a function of the scaling variable $x$ 
(defined in Eq.~(\ref{x-scaling})):
\begin{equation}
\label{scaling-law}
\frac{\varepsilon^{\prime}_{\mathrm{int}}(\beta,n_{\mathrm{ext}})}{\varepsilon^{\prime}_{\mathrm{ext}}} =
\kappa(\beta) x^\alpha \,.
\end{equation}
For the perturbative tail of 
$\varepsilon^{\prime}_{\mathrm{int}}(\beta,n_{\mathrm{ext}})/\varepsilon^{\prime}_{\mathrm{ext}}$
we keep the value of the ratio at $\beta=5$. On the other hand, the peak values have been extracted
by fitting the values around the peak to (see Fig.~4)
\begin{equation}
\label{peak-fit}
\frac{\varepsilon^{\prime}_{\mathrm{int}}(\beta,n_{\mathrm{ext}})}{\varepsilon^{\prime}_{\mathrm{ext}}} =
\frac{a_1}{a_2(\beta - \beta_{\mathrm{peak}})^2 +1} \,.
\end{equation}
From Eq.~(\ref{peak-fit}) we extract the peak value, $a_1$, and the peak position $\beta_{\mathrm{peak}}$.
It has been found that the data are compatible with the scaling law Eq.~(\ref{scaling-law}) with 
$\alpha=1.5$ (see Fig.~5). 
It is remarkable that the same power-law arises if we adopt the alternative boundary
conditions given by Eq.~(\ref{newpbcs}). So that we see that both boundary conditions 
Eq.~(\ref{pbcs}) or Eq.~(\ref{newpbcs}) lead to the same thermodynamic limit.

If we, further, take into account the shifts $\Delta \beta$ of the peak values, that
turns out to depend only on  $n_{\mathrm{ext}}$, we are led to the universal scaling-law
\begin{equation}
\label{universal}
x^{-\alpha}  
\frac{\varepsilon^{\prime}_{\mathrm{int}}(\tilde{\beta},n_{\mathrm{ext}},L_{\mathrm{eff}})}
{\varepsilon^{\prime}_{\mathrm{ext}}}  = \kappa(\tilde{\beta}) \,,
\end{equation}
where $\tilde{\beta} = \beta - \Delta \beta$. 
Indeed Figure~6 shows that all our numerical data
(for all the values of $a_H$ and $L_{\mathrm{eff}}$) can be approximately arranged on the scaling
curve $\kappa(\beta)$. Remarkably we find that the peak in $\kappa(\beta)$ is located
at $\beta_c = 2.2209(68)$ which agrees with the peak position of the specific heat extrapolated 
to the infinite volume limit $\beta_c = 2.23(2)$~\cite{Engels97}.
By using Eq.(\ref{universal}) we can determine the infinite volume limit of the 
vacuum energy density $\varepsilon_{\mathrm{int}}$. We have
\begin{eqnarray}
\label{infinite-vol}
\lefteqn{
\lim_{L_{\mathrm{eff}} \to \infty} 
\varepsilon_{\mathrm{int}}(\beta,n_{\mathrm{ext}},L_{\mathrm{eff}}) =
\varepsilon^{\prime}_{\mathrm{ext}} \, \int_0^{\tilde{\beta}} d{\tilde{\beta^{\prime}}} \,
\kappa(\tilde{\beta^{\prime}}) 
} 
\nonumber \\ 
& & \qquad \qquad \qquad \qquad \qquad \qquad 
\times  \lim_{L_{\mathrm{eff}} \to \infty} \left( \frac{a_H}{L_{\mathrm{eff}}} \right)^\alpha
\nonumber \\
& & \simeq \frac{H^2}{2 \beta} \left[ \int_0^{\tilde{\beta}} d{\tilde{\beta^{\prime}}} \, 
\kappa(\tilde{\beta^{\prime}}) \right] \times 
\lim_{L_{\mathrm{eff}} \to \infty} \left( \frac{a_H}{L_{\mathrm{eff}}} \right)^\alpha = 0 \,,
\end{eqnarray} 
in the whole range of $\beta$. This in turn implies that in the continuum limit 
$(L_{\mathrm{eff}} \to \infty, \beta \to \infty)$ the SU(2) vacuum completely screens
the external chromomagnetic Abelian field. In other words, the continuum vacuum behaves 
as an Abelian magnetic condensate medium in accordance with the dual superconductivity scenario. 

\renewcommand{\thesection}{\normalsize{\Roman{section}.}}
\section{\normalsize{MONTE CARLO SIMULATIONS: $T \ne 0$}}
\renewcommand{\thesection}{\arabic{section}}

We can extend the study of the SU(2) gauge system in an external chromomagnetic Abelian field 
to the case of finite temperature. As it is well known the relevant dynamical quantity is 
the free energy. On the lattice the physical temperature $T_{\mathrm{phys}}$ is introduced
by (in units of $\kappa_B =1$):
\begin{equation}
\label{T-phys}
\frac{1}{T_{\mathrm{phys}}} = L_t \cdot a \,,
\end{equation}
where $L_t$ is the linear extension in the time direction $L_t=L_4$, while the
extension on the spatial direction should be infinite. In numerical simulations, however, 
the spatial extension would of  course be finite. In order to approximate the thermodynamic
limit one should respect the relation 
\begin{equation}
\label{Ls-gg-Lt}
L_s \gg L_t   \,.
\end{equation}
We perform our numerical simulation on $32^3 \times L_t$ lattices by imposing 
\begin{equation}
\label{LtsuLs-le-4}
\frac{L_t}{L_s} \le 4 
\end{equation}
in order to avoid finite volume effects.

In the case of constant external chromomagnetic field the relevant quantity is the density
of the free energy
\begin{equation}
\label{free-energy}
F[\vec{A}^{\mathrm{ext}}] = - \frac{1}{V L_t} 
\ln \frac{\mathcal{Z}[\vec{A}^{\mathrm{ext}}]}{{\mathcal{Z}}[0]} \,, \quad V=L_s^3 \,.
\end{equation}
The pure gauge system undergoes the deconfinement phase transition by increasing the 
temperature. The order parameter for the deconfinement phase transition is the
Polyakov loop
\begin{equation}
\label{Polyakov-loop}
P = {\mathrm{Tr}} \prod_{x_4=1}^{L_t} U_4(x) \,.
\end{equation}
As a preliminary step we look at the behavior of the temporal Polyakov loop $\langle P \rangle$
versus the external applied field. We start with the SU(2) gauge system at $\beta=2.5$ on
$32^3 \times 5$ lattice at zero applied external field (i.e. $n_{\mathrm{ext}}=0$) that is 
known to be in the deconfined phase of finite temperature SU(2). 
If the external field strength is increased the expectation value of the Polyakov loop is driven 
towards the value at zero temperature (see Fig.~7). Similar behavior has been reported 
by the authors of Ref.~\cite{Ogilvie} within a different approach. It is worthwhile to
stress that our result is consistent with the dual superconductor mechanism of confinement.

On the other hand, if we start with the SU(2) gauge system at zero temperature in a 
constant Abelian chromomagnetic background field of fixed strength ($n_{\mathrm{ext}}=1$)
and increase the temperature, then we find that the perturbative tail of the $\beta$-derivative
of the free energy density $F^{\prime}_{\mathrm{int}}(\beta,n_{\mathrm{ext}})/
\varepsilon^{\prime}_{\mathrm{ext}}$ increases with $1/L_t$ and tends towards
the ``classical'' value $F^{\prime}_{\mathrm{int}}(\beta,n_{\mathrm{ext}})/
\varepsilon^{\prime}_{\mathrm{ext}} \simeq 1 $ (see Fig.~8).

We may  conclude, then, that by increasing the temperature there is no screening effect
in the free energy density confirming that the zero-temperature screening of the 
external field is related to the confinement.
Moreover the information of   $F^{\prime}_{\mathrm{int}}(\beta,n_{\mathrm{ext}})/
\varepsilon^{\prime}_{\mathrm{ext}}$ at finite temperature can be used to get an estimate 
of the deconfinement temperature $T_c$. In Figure 9 we magnify the peak region for various values
of $L_t$. We see clearly that the pseudocritical coupling $\beta^*(L_t)$ depends on $L_t$.
To determine the pseudocritical couplings we parametrize $F^{\prime}_{\mathrm{int}}(\beta,L_t)$
near the peak as
\begin{equation}
\label{peak-form}
\frac{F^{\prime}_{\mathrm{int}}(\beta,L_t)}{\varepsilon^{\prime}_{\mathrm{ext}}} =
\frac{a_1(L_t)}{a_2(L_t) [\beta - \beta^*(L_t)]^2 +1} \,.
\end{equation}
We restrict the region near $\beta^*(L_t)$ until the fits Eq.~(\ref{peak-form}) give a reduced
$\chi^2$ of order $1$.

Having determined $\beta^*(L_t)$ we estimate the deconfinement temperature as 
\begin{equation}
\label{Tc}
\frac{T_c}{\Lambda_{\mathrm{latt}}} = \frac{1}{L_t-1} \frac{1}{f(\beta^*(L_t))} \,,
\end{equation}
where
\begin{equation}
\label{af}
f(\beta) = \left( \frac{11}{6 \pi^2} \frac{1}{\beta} \right)^{-51/121} \, 
\exp(-\frac{3 \pi^2}{11} \beta) \,.
\end{equation}
In Eq.(~\ref{Tc}) we take into account that, due to the frozen time slice,
the effective extension in the time direction is $L_t^{\mathrm{eff}}=L_t-1$.

In Figure 10 we display $T_c/\Lambda_{\mathrm{latt}}$ for different temperatures.
Following Ref.~\cite{Fingberg1993} we perform a linear extrapolation to the continuum of our data
for $T_c/\Lambda_{\mathrm{latt}}$.
We see that our estimate of $T_c/\Lambda_{\mathrm{latt}}$ in the continuum
is in fair agreement with the one available
in the literature~\cite{Fingberg1993}:
\begin{equation}
\label{Tc-Fingberg}
\frac{T_c}{\Lambda_{\mathrm{latt}}} = 24.38 \pm 2.18 \,.
\end{equation}

\renewcommand{\thesection}{\normalsize{\Roman{section}.}}
\section{\normalsize{CONCLUSIONS}}
\renewcommand{\thesection}{\arabic{section}}

We have studied the non-perturbative dynamics of the vacuum of SU(2) lattice gauge theory 
by means of the gauge-invariant effective action defined using the lattice Schr\"odinger functional.

At zero temperature our numerical results indicate that in the continuum limit 
$L_{\mathrm{eff}} \to \infty$, $\beta \to \infty$ we have
\begin{equation}
\label{climit}
\varepsilon[ \vec{A}^{\mathrm{ext}} ] = 0 \,,
\end{equation}
so that the vacuum screens completely the external chromomagnetic Abelian field. 
In other words, the continuum vacuum behaves as an Abelian magnetic condensate medium
in accordance with the dual superconductivity scenario. In particular we have
\begin{equation}
\label{HB}
\varepsilon[ \vec{A}^{\mathrm{ext}} ] \sim H^a_k B^a_k = \frac{1}{\mu} F^a_{ij}  F^a_{ij} \,
\end{equation}
where $\mu$ is the vacuum color magnetic permeability. Thus Eq.~(\ref{climit}) implies that
$\mu \to \infty$ in the continuum limit. As a consequence by Lorentz invariance the
vacuum color dielectric constant tends to zero. This in turns implies that the vacuum does not
support an isolated color charge, i.e. the color confinement.

The intimate connection between the screening of the external background field and the
confinement is corroborated by the finite temperature results. Indeed our numerical data 
show that the zero-temperature screening of the external field is removed by increasing
the temperature. Moreover, at finite temperature it seems that confinement is restored 
by increasing the strength of the external applied field. 

At finite temperature we find that the $\beta$-derivative of the free energy density behaves
like a specific heat. From the peak position of the $\beta$-derivative of the free energy density
we obtained an estimation of the critical temperature $T_c/\Lambda_{\mathrm{latt}}$ that extrapolates
in the continuum limit to a value consistent with previous determinations in the literature.

Let us  conclude by stressing that our method can be easily extended to the SU(3) gauge theory.
Moreover we also feel that the lattice gauge invariant effective action could be also employed
to study different background fields.

%

\begin{figure}[H]
\label{Fig1}
\begin{center}
\includegraphics[clip,width=0.85\textwidth]{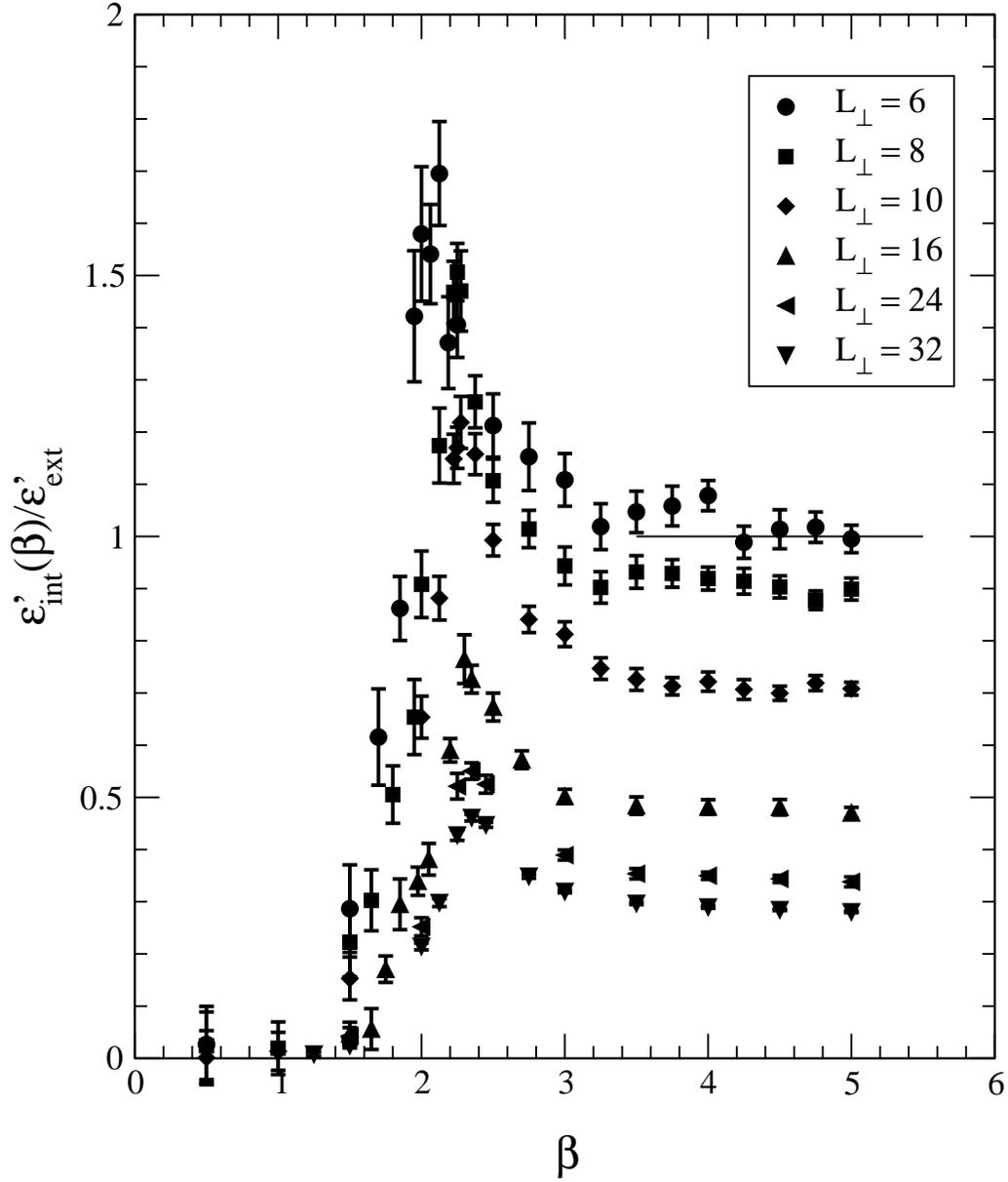}
\caption{The $\beta$-derivative of the 
internal energy density Eq.~(\ref{deriv}) versus $\beta$ at different 
values of the transverse lattice size $L_\perp$.}
\end{center}
\end{figure}
\begin{figure}[H]
\label{Fig2}
\begin{center}
\includegraphics[clip,width=0.85\textwidth]{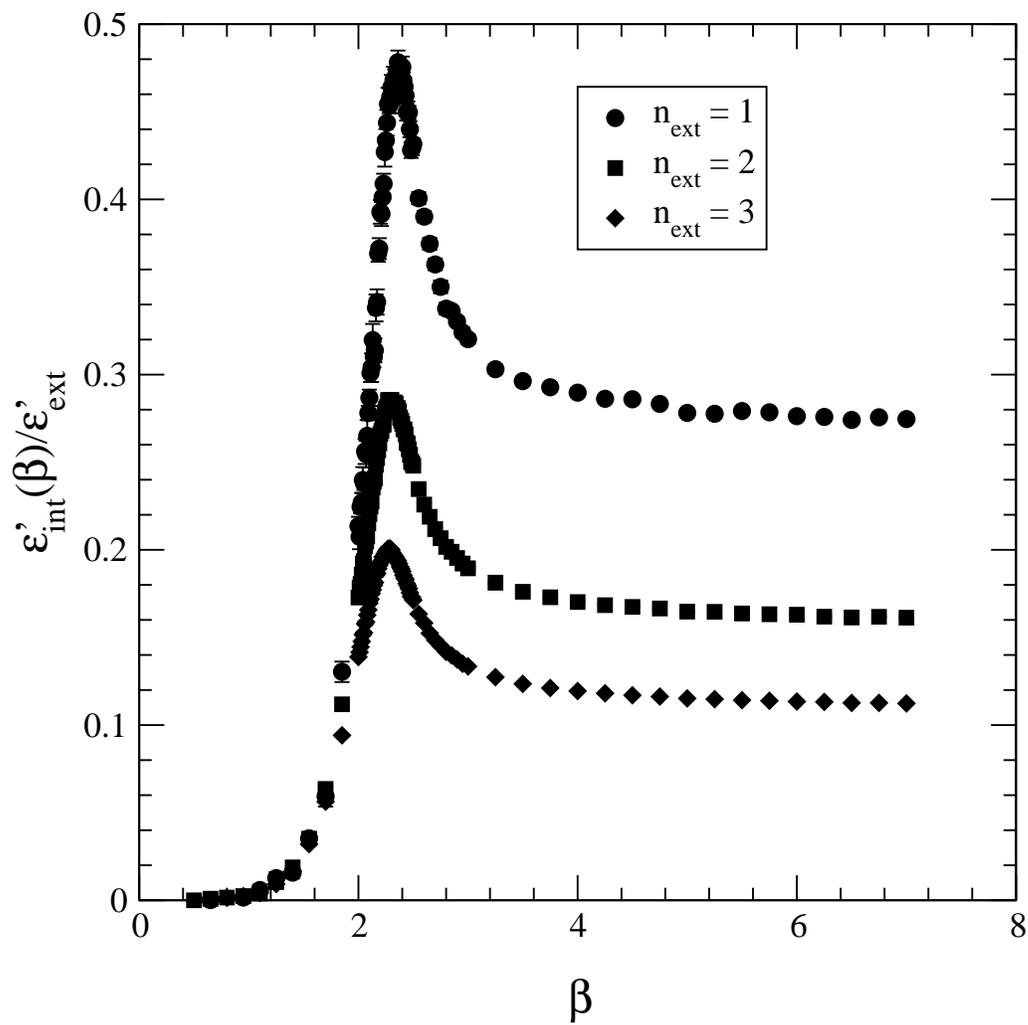}
\caption{The $\beta$-derivative of the internal energy density Eq.~(\ref{deriv}) versus $\beta$ for a 
transverse lattice size $L_\perp=32$ at different values of applied external 
field strength.}
\end{center}
\end{figure}
\begin{figure}[H]
\label{Fig3}
\begin{center}
\includegraphics[clip,width=0.85\textwidth]{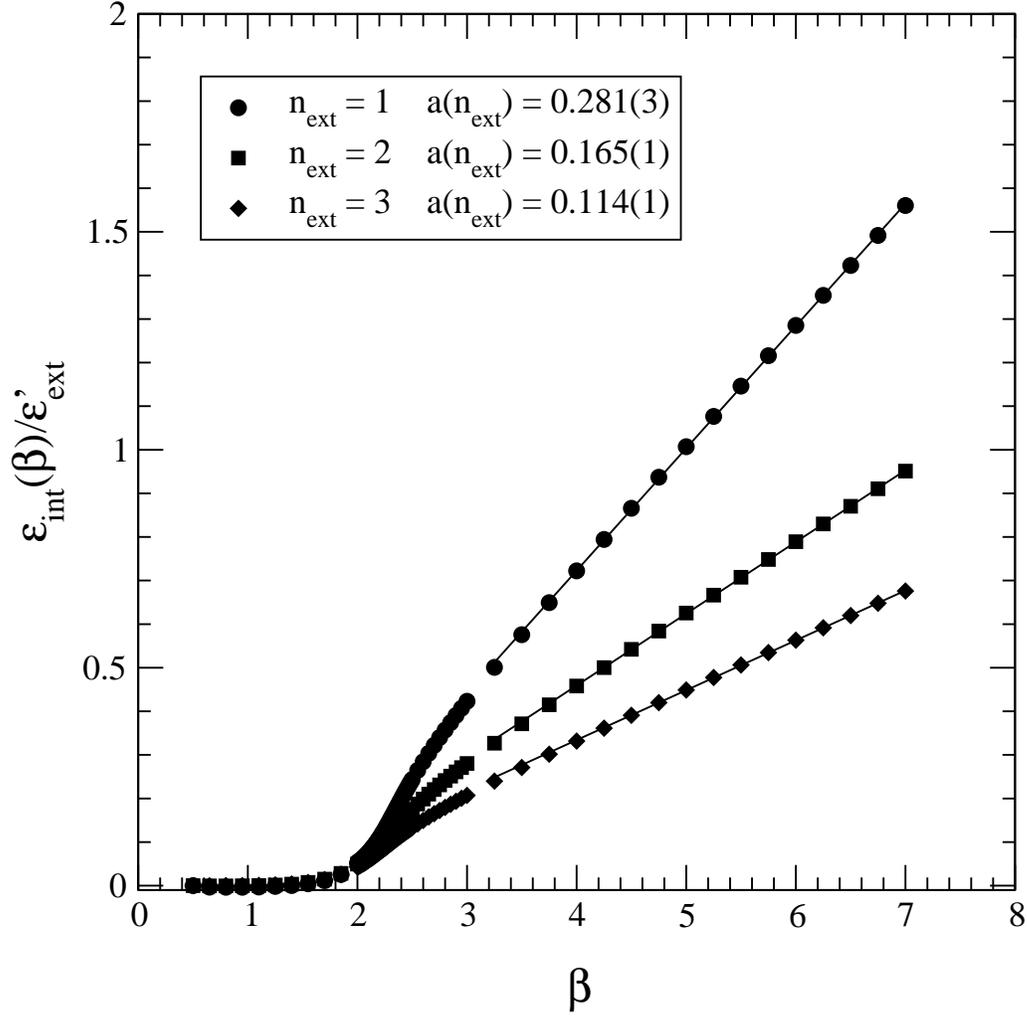}
\caption{The  energy density Eq.~(\ref{trapez}) versus $\beta$ for  
transverse lattice size $L_\perp=32$ at different values of applied external 
field strength. The solid lines are the linear fits Eq.~(\ref{betagg1}).}
\end{center}
\end{figure}
\begin{figure}[H]
\label{Fig4}
\begin{center}
\includegraphics[clip,width=0.85\textwidth]{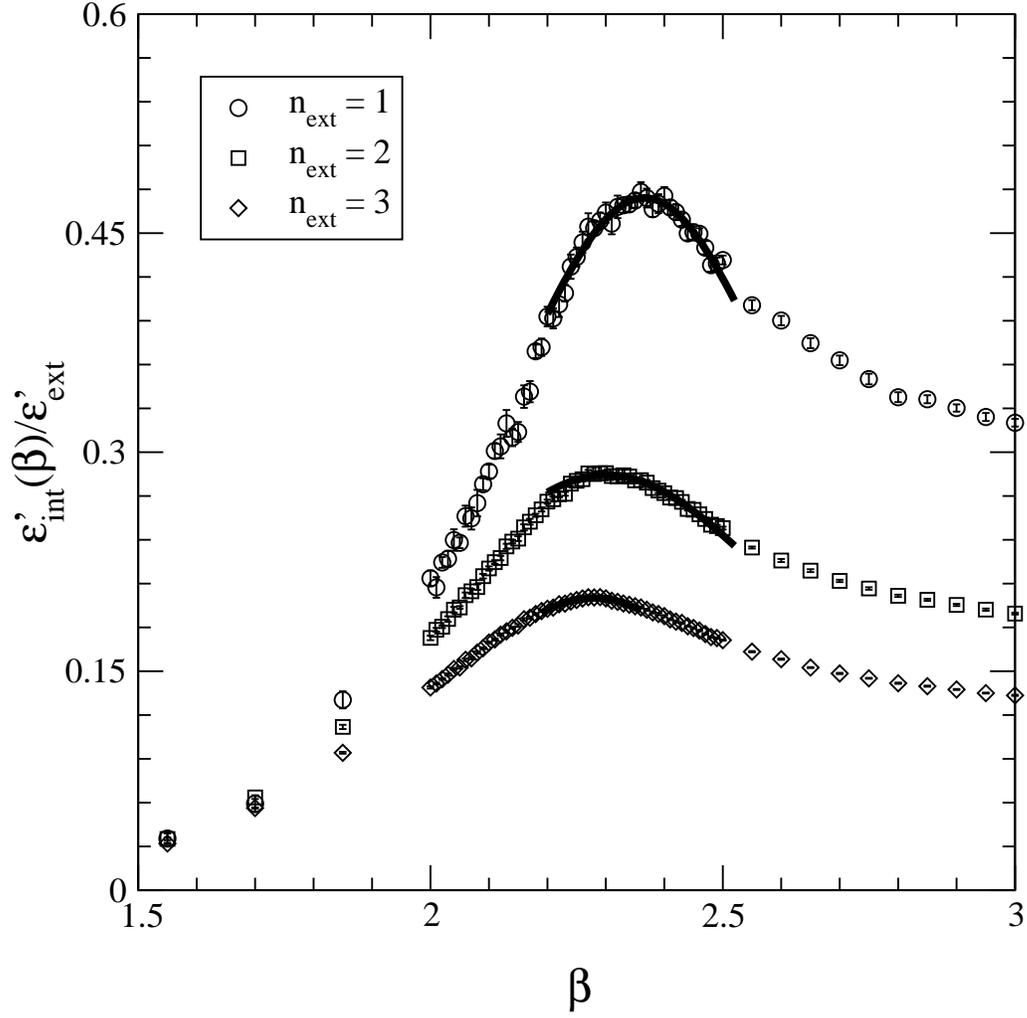}
\caption{The data displayed in Fig.~2 near the peaks of the $\beta$-derivative of the 
internal energy density Eq.~(\ref{deriv}) in correspondence of each value
of $n_{\mathrm{ext}}$. The solid lines are the fits Eq.~(\ref{peak-fit}).}
\end{center}
\end{figure}
\begin{figure}[H]
\label{Fig5}
\begin{center}
\includegraphics[clip,width=0.85\textwidth]{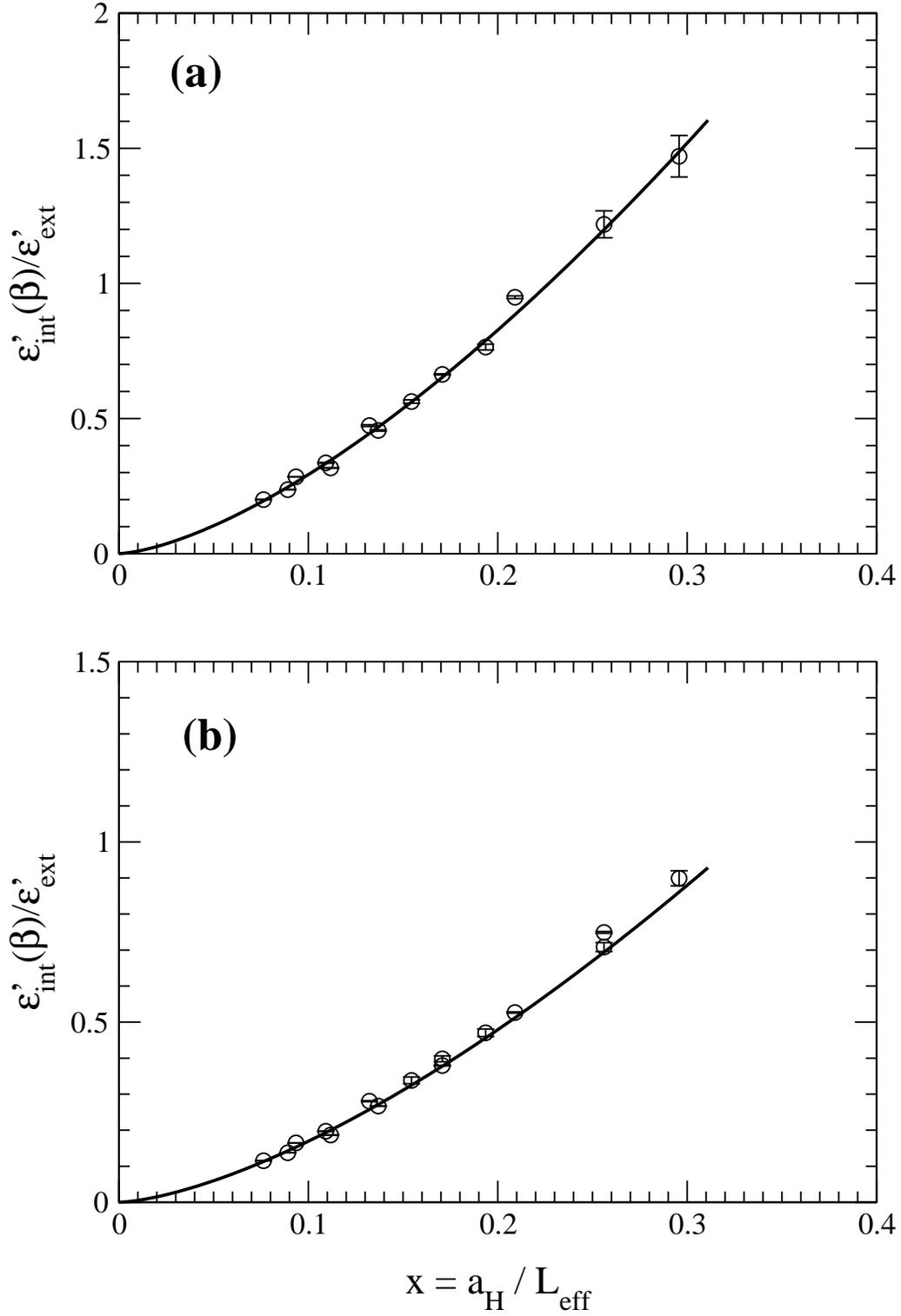}
\caption{The lattice data for 
$\varepsilon^{\prime}_{\mathrm{int}}(\beta,n_{\mathrm{ext}})/\varepsilon^{\prime}_{\mathrm{ext}}$ at
different values of the transverse lattice size $L_\perp$ and external field strength $n_{\mathrm{ext}}$
for $\beta=5$ (a) and $\beta=\beta_{\mathrm{peak}}$ (b), 
versus the scaling variable defined in Eq.~(\ref{x-scaling}). The solid lines are
the fits Eq.~(\ref{scaling-law}) with $\alpha=1.5$.}
\end{center}
\end{figure}
\begin{figure}[H]
\label{Fig6}
\begin{center}
\includegraphics[clip,width=0.85\textwidth]{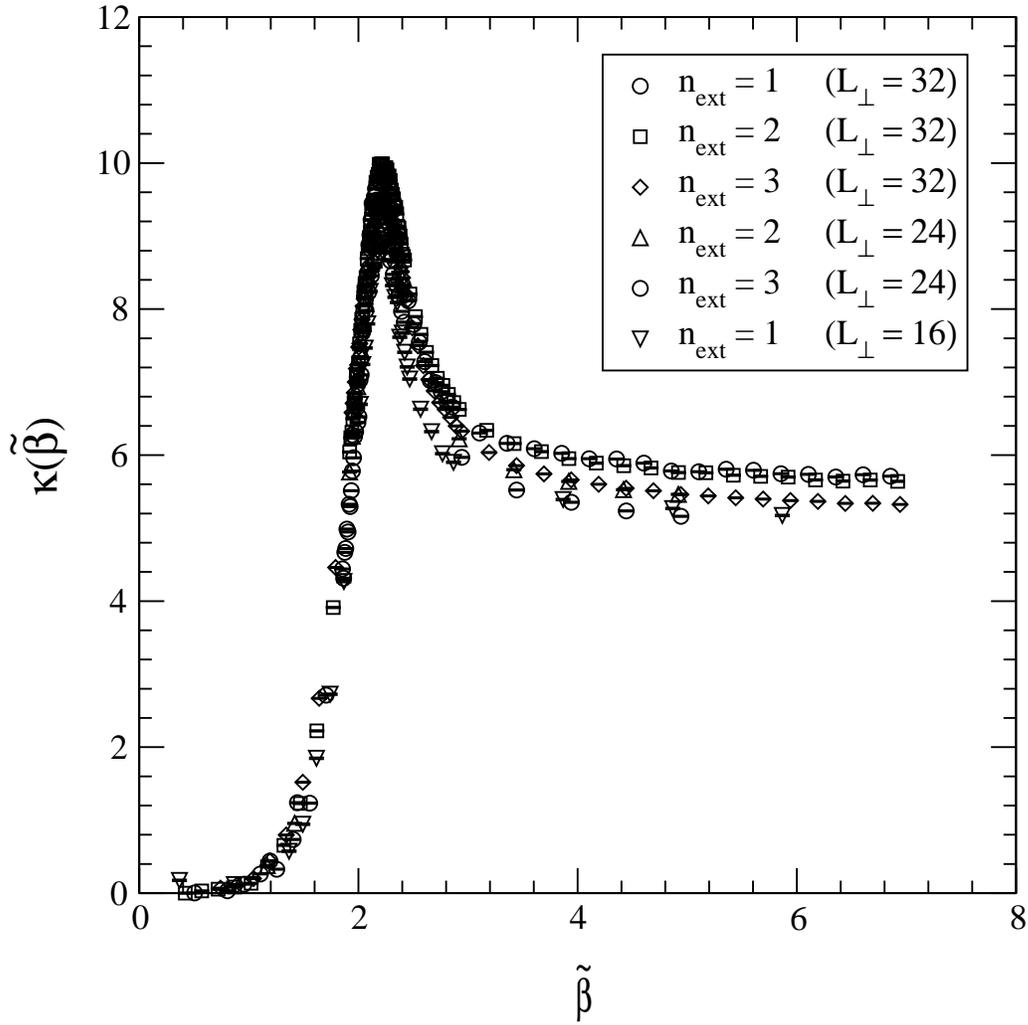}
\caption{The scaling curve obtained by re-scaling all lattice data for 
$\varepsilon^{\prime}_{\mathrm{int}}(\tilde{\beta},n_{\mathrm{ext}},L_{\mathrm{eff}})$
according to Eq.~(\ref{universal}).} 
\end{center}
\end{figure}
\begin{figure}[H]
\label{Fig7}
\begin{center}
\includegraphics[clip,width=0.85\textwidth]{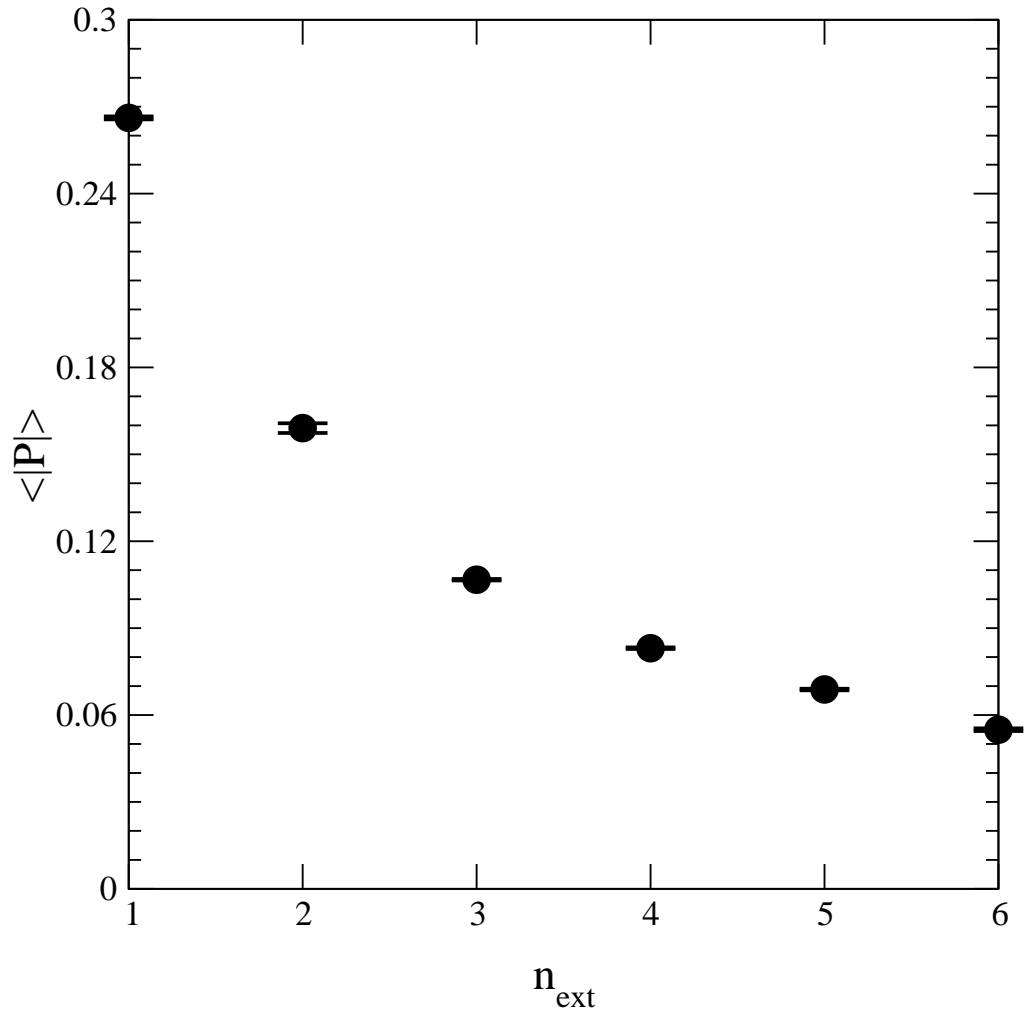}
\caption{The absolute value of the Polyakov loop on a $32^3\times5$ lattice at
$\beta=2.5$ versus $n_{\mathrm{ext}}$.}
\end{center}
\end{figure}
\begin{figure}[H]
\label{Fig8}
\begin{center}
\includegraphics[clip,width=0.85\textwidth]{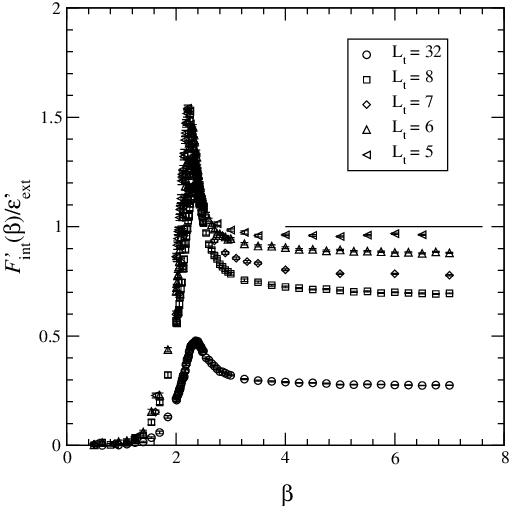}
\caption{The $\beta$-derivative
of the free energy density Eq.~(\ref{free-energy}) 
$F^{\prime}_{\mathrm{int}}(\beta,n_{\mathrm{ext}})/
\varepsilon^{\prime}_{\mathrm{ext}}$ versus $\beta$ at different values
of the temporal lattice extension $L_t$.}
\end{center}
\end{figure}
\begin{figure}[H]
\label{Fig9}
\begin{center}
\includegraphics[clip,width=0.85\textwidth]{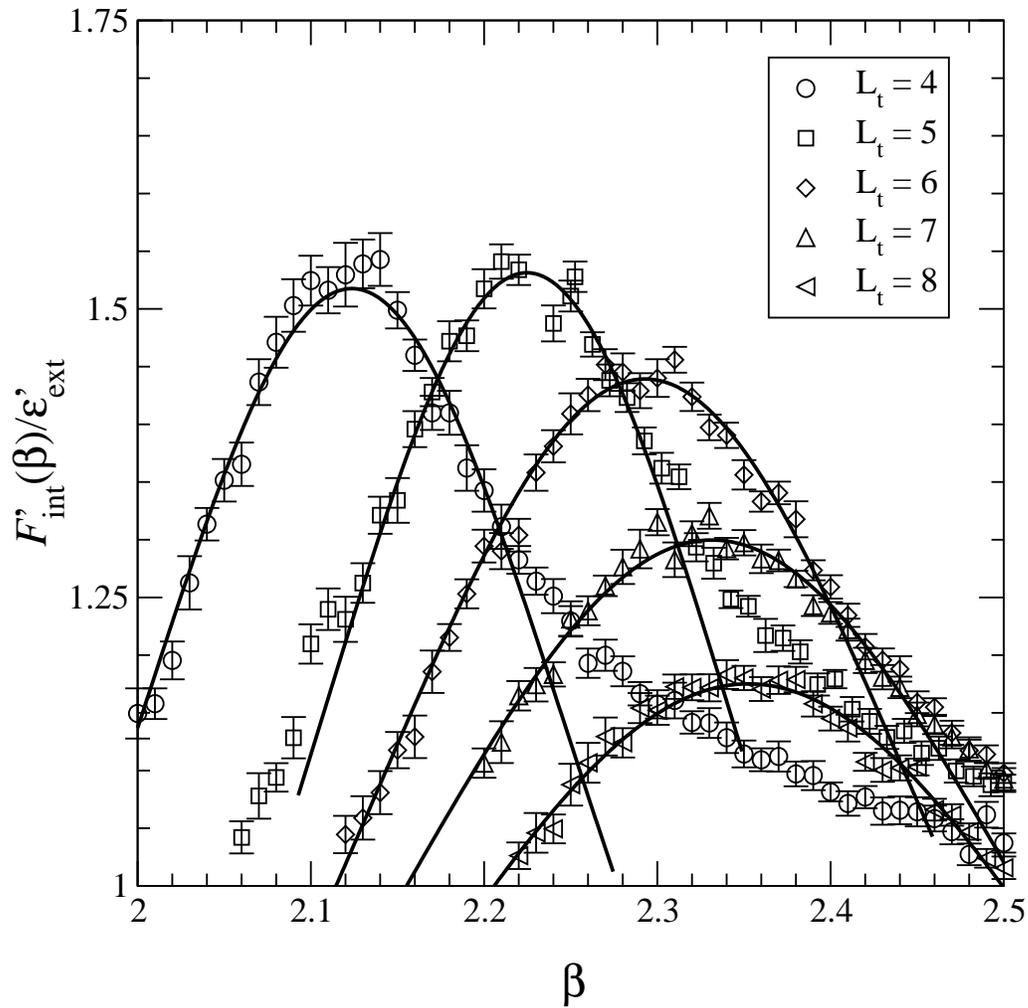}
\caption{The peak region for various values of $L_t$. The solid lines are the fits
Eq.~(\ref{peak-form}).}
\end{center}
\end{figure}
\begin{figure}[H]
\label{Fig10}
\begin{center}
\includegraphics[clip,width=0.85\textwidth]{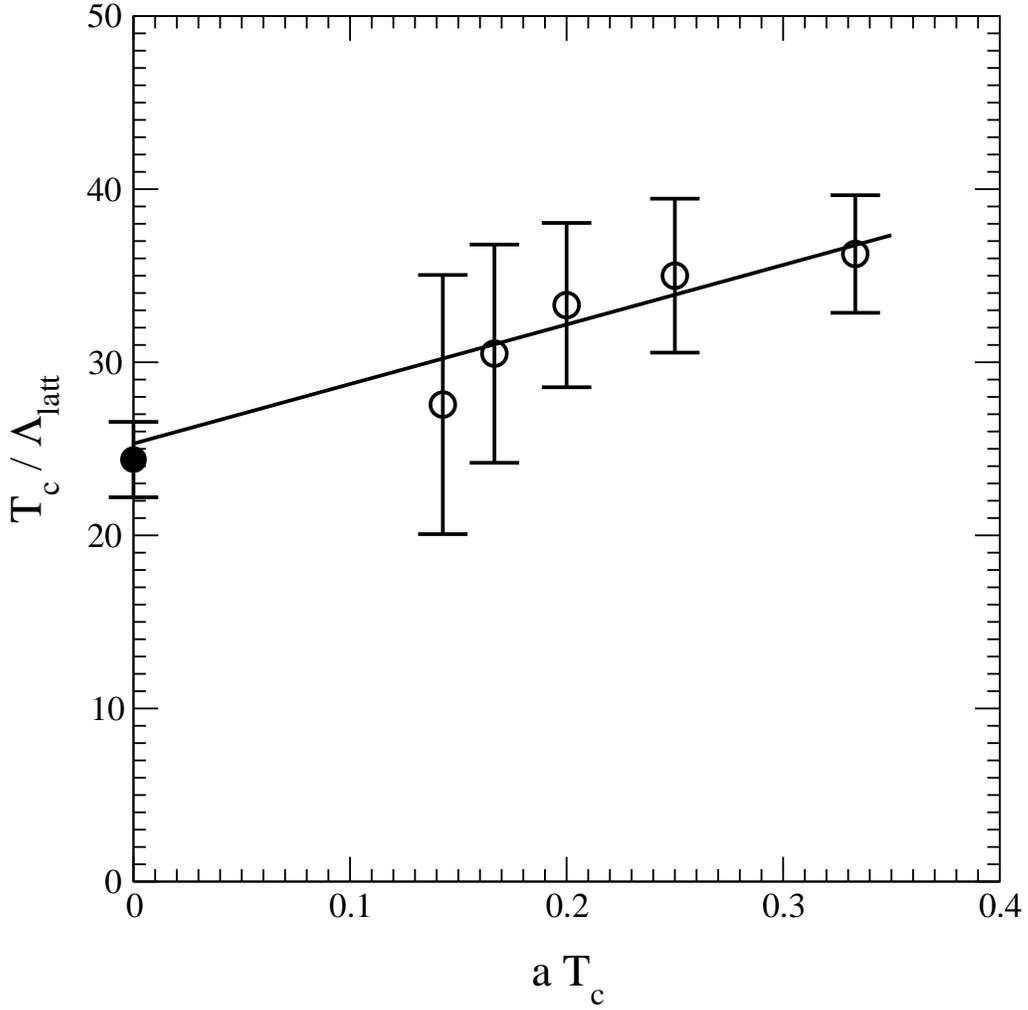}
\caption{Our lattice data for $T_c/\Lambda_{\mathrm{latt}}$ versus the temperature (circles). 
The full circle is the continuum estrapolation of Ref.~\protect\cite{Fingberg1993}
(see Eq.~(\ref{Tc-Fingberg})). The solid line is a linear fit to our data.}
\end{center}
\end{figure}

\end{document}